\documentclass[floatfix,twocolumn,twoside,preprintnumbers,amsmath,amssymb,showkeys]{revtex4}
\usepackage{epsfig}
\usepackage{graphicx}

\usepackage{axodraw}

\usepackage{fancyhdr}

\usepackage{pslatex}

\begin{document}

\title{Fluctuations and Fermi-Dirac Correlations in $e^+e^-$-annihilation
\footnote{Talk presented at XXXVI Int. Symp. on Multiparticle Dynamics,
Paraty, Brazil, 2 - 8 Sept. 2006.}}

\author{G\"osta Gustafson}

\affiliation{Lund University, Dept. of Theoretical Physics, 
S\"olvegatan 14A, S-223 62 Lund, Sweden\\Gosta.Gustafson@thep.lu.se}


\begin{abstract}

In this talk I first present a short review of fluctuations in 
$e^+e^-$-annihilations. I then describe some new results on FD correlations.
Experimental analyses of $pp$ and $\Lambda\Lambda$ correlations
indicate a very small production radius. This result relies
very strongly on comparisons with MC simulations. A study of the
approximations and uncertainties is these simulations imply that
it is premature to draw such a conclusion from the data.


\end{abstract}
\maketitle

\thispagestyle{fancy}

\setcounter{page}{1}


\section{Fluctuations}

In bremsstrahlung in QED, the emission of a photon does not change 
the current for subsequent emissions. This implies that the photon
multiplicity is described by a Poisson distribution. As the gluons carry
colour charge, the emission of a gluon in QCD changes the current
relevant for the subsequent emissions.
An initial hard gluon will radiate many softer gluons.
This leads to a cascade with an exponential growth in multiplicity
and to large fluctuations. The multiplicity distribution is 
approximately satisfying KNO scaling, with a width which is
proportional to the average multiplicity.
This short review is divided in three parts: 
\vspace{1mm}

A. Multiplicity distribution for partons

B. Effects of hadronization

C. Fractal structures
\vspace{-2mm}
\subsection{Multiplicity distribution for partons}

\begin{figure}
\includegraphics*[bb=220 608 510 770, width=8cm]{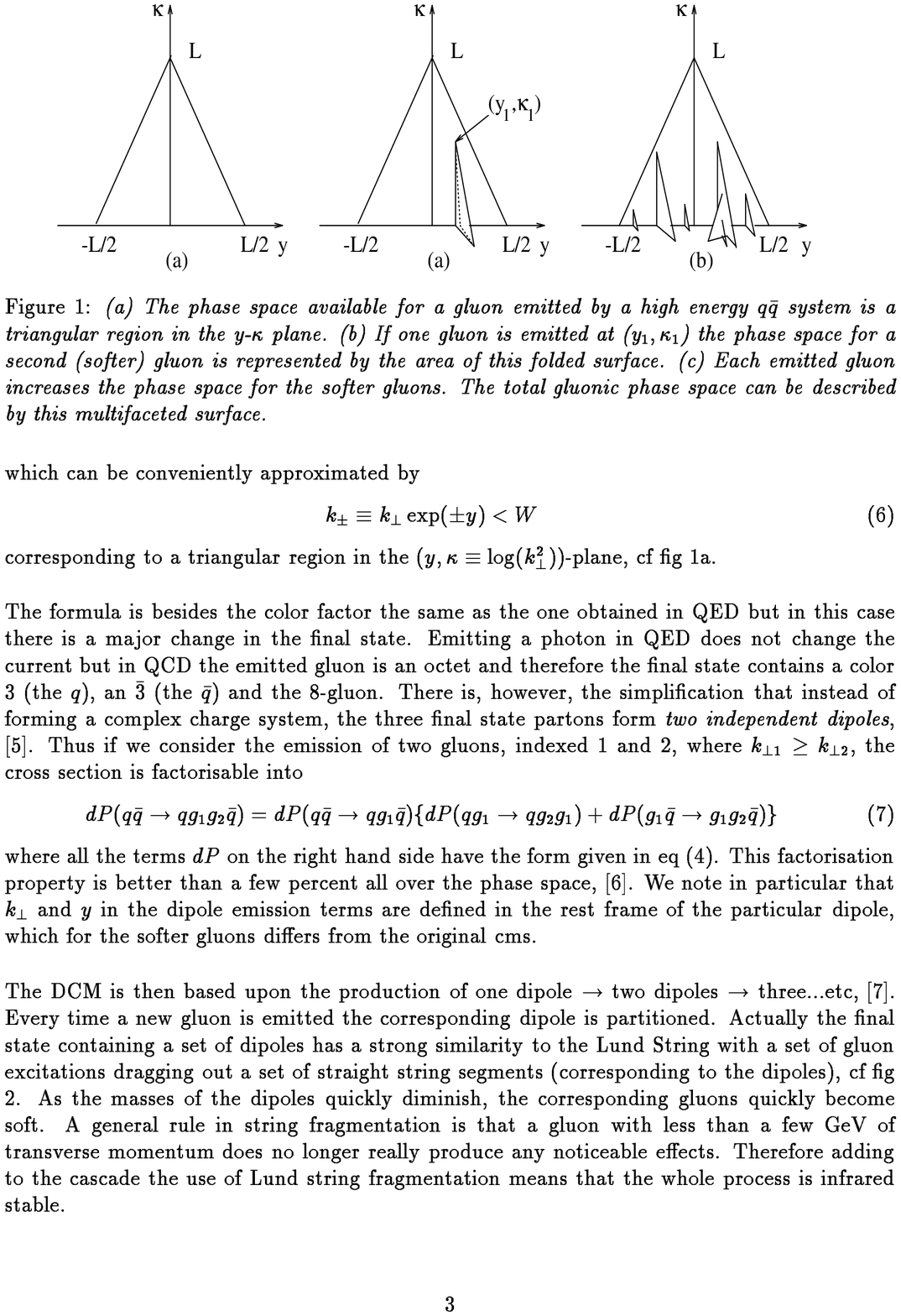}
\caption{\label{fig:phasespace}
The phase space for gluon emission in $e^+e^-$-ann. is a triangular 
region in the 
$(y,\kappa=\ln k_\perp^2)$-plane. The height of the triangle 
is given by $L=\ln s$. When one gluon is emitted at $(y_1, \kappa_1)$
a second gluon with smaller $k_\perp$ can be emitted in the larger phase 
space shown in fig.~(a).
Further softer emissions give the fractal phase space in fig.~(b).}
\end{figure}

In $e^+e^-$-ann. the emission of a gluon from a $q\bar{q}$
pair is a \emph{coherent} emission from e.g. a \emph{red} and an
\emph{anti-red} colour charge. The result corresponds to a colour
dipole with the following distribution (in the leading log approximation):
\begin{equation}
d N = \bar{\alpha}(k_\perp^2) \frac{d k_\perp^2}{k_\perp^2} d y,
\,\,\,\,\mathrm{with}\,\, \bar{\alpha} = \frac{3 \alpha_s}{2\pi}.
\label{eq:dipole}
\end{equation}
The phase space is a triangular region in $(y,\ln k_\perp^2)$-space
given by $|y|<\ln(s/k_\perp^2)/2$, and within this triangle
the density is just given by $\bar{\alpha}$. 
The emission of a second (softer) gluon is given by two dipoles,
one between the quark and the first gluon, and one between this
gluon and the antiquark. The phase space is enlarged compared
to that for the first gluon, and corresponds to the folded surface shown in 
fig.~\ref{fig:phasespace}a. For subsequent gluons the phase space
is further increased, and corresponds to the fractal surface in
fig.~\ref{fig:phasespace}b. 

We let $P(N, L=\ln s)$ denote the distribution in the number, $N$, of dipoles
(which neglecting $q\bar{q}$ pair creation is equal to the number of 
gluons plus one).
The Laplace transform $\mathcal{P}(\gamma,L)$ is given by 
\begin{equation}
\mathcal{P}(\gamma,L) \equiv \sum_N e^{-\gamma N} P(N,L).
\label{eq:mathcalP}
\end{equation}
It is then relatively easy to derive the relation (see e.g. 
ref.~\cite{Andersson:1988ee})
\begin{equation}
\frac{d^2 \ln \mathcal{P}}{d L^2} = \frac{\alpha_0}{L} (\mathcal{P} - 1),
\,\,\,\,\,\,\mathrm{where}\,\,\, \bar{\alpha} \equiv \frac{\alpha_0}{L}.
\label{eq:evolution}
\end{equation}
Expanding eq.~(\ref{eq:mathcalP}) in a Taylor series in $\gamma$ gives
\begin{equation}
\mathcal{P}=1-\gamma \langle N\rangle + {\small \frac{1}{2}}\, \gamma^2 
\langle N^2\rangle + \ldots
\label{eq:Pexpansion}
\end{equation}
Inserted in eq.~(\ref{eq:evolution}) this also implies the following equations
for the average and the variance of the multiplicity distribution:
\begin{eqnarray}
\frac{d^2 }{d L^2} \langle N\rangle &=& \frac{\alpha_0}{L} \langle N\rangle \label{eq:eqN} \\
\frac{d^2 }{d L^2} \,\,(\langle N^2\rangle\!&-&\!\langle N\rangle^2) = \frac{\alpha_0}{L} 
\langle N^2\rangle.
\label{eq:eqN2}
\end{eqnarray}
The solutions to these equations are Bessel functions, and for high energies
we get
\begin{eqnarray}
&\langle N\rangle\!\!& \sim \,L^{1/4} \exp(2\sqrt{\alpha_0 L})  \nonumber \\
&V\!\!\!\! & \equiv \, \langle N^2\rangle\!-\!\langle N\rangle^2 \,\approx \frac{1}{3} 
\langle N\rangle^2.
\label{eq:solN}
\end{eqnarray}
We see that the width of the distribution is proportional to 
$\langle N\rangle$, in 
agreement with KNO scaling, and thus for large $N$ the distribution is much wider 
than a Poisson distribution. 

The anomalous dimension is given by a logarithmic derivative. Note that
in the literature
one can find different definitions, where the derivative is taken 
with respect either to $\ln s$ or to $\ln W$:
\begin{eqnarray}
\gamma_0^{(s)} & \equiv & \frac{d \ln \langle N\rangle}{d \ln s} = \sqrt{\bar{\alpha}}=
\sqrt{\frac{3\alpha_s}{2\pi}}, \\
\gamma_0^{(W)} & \equiv & \frac{d \ln \langle N\rangle}{d \ln W} = 2\sqrt{\bar{\alpha}}=
\sqrt{\frac{6\alpha_s}{\pi}}.
\label{eq:gamma0}
\end{eqnarray}

I want here to add a few comments:

\begin{itemize}
\item $\langle N\rangle$ depends on the resolution, i.e. on the cut-off for soft 
emissions. It is here essential to have a cut-off which is defined
locally, and not fixed in e.g. the overall cms. In the dipole cascade 
model this is chosen as $k_\perp$ measured 
in the rest frame of the emitting dipole.
\item The running of $\alpha_s$ is important. With a constant $\alpha_s$
the multiplicity would grow proportional to $\exp(\sqrt{\bar{\alpha}}\cdot L)$.
i.e. much faster than the result in eq.~(\ref{eq:eqN}).
\item Non-leading corrections are \emph{very large}. Terms suppressed by 
factors $1/\sqrt{L}$ in the evolution \emph{equation} give extra powers of $L$
as factors in the \emph{solution}. Also including NLL terms, the result
is sensitive to effects of still higher order. For a more detailed discussion 
see e.g. ref.~\cite{Gustafson:1993dd}.
\end{itemize}
\vspace{-2mm}
\subsection{Hadronization effects}

In string fragmentation the average hadron multiplicity, $\langle n\rangle$, 
in a single $q\bar{q}$
system is proportional to $\ln(s/s_0)$, where $s_0$ is a scale of order 1GeV.
The variance of the fluctuations, $V=\langle n^2\rangle-\langle n\rangle^2$, is proportional to 
$\langle n\rangle$, and the distribution is thus relatively more narrow for higher 
energies and larger $\langle n\rangle$. For a system of a $q\bar{q}$ pair and a number of
gluons we get a set of string pieces with (squared) masses 
$s_{i,i+1}=(q_i + q_{i+1})^2$, where $q_i$ is the momentum of parton $i$ 
and the partons are ordered in colour. For the average multiplicity we then 
get
\begin{equation}
\langle n\rangle \propto \sum \ln(s_{i,i+1}/s_0)
\label{eq:lambda}
\end{equation}
which just corresponds to the length of the baseline of the fractal surface in 
fig.~\ref{fig:phasespace}b.

In eq.~(\ref{eq:lambda}) the mass of a string piece should not be 
smaller than $\sqrt{s_0}$, as it otherwise would give a negative contribution. 
It is, however, also possible to define an infrared stable measure, called
the $\lambda$-measure,
which is insensitive to the cut-off for the perturbative cascade
\cite{Andersson:1989ww}. Simulations show that a high energy even with little
gluon radiation and an event at lower energy but more radiation give equally many
hadrons if they have the same $\lambda$-value.
Thus the multiplicity depends only on the measure $\lambda$, which
corresponds to an ``effective string length''.
The probability density $P(\lambda,L)$ satisfies the same evolution equation
(\ref{eq:evolution}) as $P(N,L)$, only with different boundary conditions.
This implies that the distribution in $\lambda$ has the same high
energy behaviour as the distribution in dipole multiplicity, 
eq.~(\ref{eq:eqN}).

The fluctuations in the hadron multiplicity get contributions from both 
the perturbative cascade and the soft hadronization, and we get approximately
\begin{equation}
\frac{V(n,s)}{\langle n(s)\rangle^2} = \frac{V(\lambda,s)}{\langle \lambda(s)\rangle^2} +
\frac{V(n,\langle \lambda\rangle)}{\langle n(\langle \lambda\rangle)\rangle^2}. 
\label{eq:varance}
\end{equation}
Here the first term corresponds to the fluctuations in the cascade and the 
second one to those in the hadronization process. At high energy the 
first contribution dominates, while the second is larger at energies below 
40-50 GeV. In ref.~\cite{Andersson:1989ww} it is shown that the total result
satisfies KNO scaling with $V=const \cdot \langle n\rangle^2$,
although this is not the case for any of the contributions separately 
below the top LEP energy. At higher energies the hadronization contribution
can be neglected.

\subsection{Fractal structures and intermittency}

The baseline of the surface in fig.~\ref{fig:phasespace}b looks like a Koch
snowflake curve. Looking at the curve with a coarser resolution means 
that only emissions with $k_\perp > k_{\perp\, \mathrm{res}}$ are 
included. This corresponds to cutting the surface in fig.~\ref{fig:phasespace}b
at a higher level, corresponding to $\ln k_{\perp \, \mathrm{res}}^2$. 
The length of this
curve satisfies again the same evolution equation. The boundary conditions
are different, and with the notations $L=\ln s$ and $\kappa = \ln k_\perp^2$ 
the length will be proportional to the expression
\begin{equation}
Length \sim L^{1/4} \kappa_\mathrm{res}^{3/4}  
\exp(2\sqrt{\alpha_0 L} - 
2\sqrt{\alpha_0 \kappa_\mathrm{res}})
\end{equation}
It is also possible to define a (multi)fractal dimension for this curve
\begin{equation}
D = 1 - \frac{d\ln length}{d\ln(\kappa_\mathrm{res})} = 
1+ \sqrt{\frac{\alpha_0}{\kappa_\mathrm{res}}} = 1 +\gamma_0^{(s)}
\end{equation}
(It is called a multifractal as the dimension varies with the resolution 
$\kappa_\mathrm{res}$.)

Fluctuations in small phase space regions, e.g. slices in rapidity $\Delta y$,
can be described by the scaled moments
\begin{eqnarray}
C_q \equiv \frac{\langle n^q\rangle}{\langle n\rangle^q }\, ;\,\,\,\,\,\,
F_q \equiv \frac{\langle n(n-1) \ldots (n-q+1)\rangle}{\langle n\rangle^q }
\label{eq:moments}
\end{eqnarray}

If the total phase space is divided in $M$ equal pieces (which gives $\Delta y=Y/M$)
a scaling behaviour $F_q \propto M^{\phi_q}$ is interpreted as an
signal for \emph{intermittency}. This can be associated with a Renyi 
(multifractal) dimension
\begin{equation}
D_q \equiv 1 - d_q \equiv 1 - \frac{\phi_q}{q-1}
\label{eq:Renyi}
\end{equation}
In the perturbative cascade these fluctuations in small bins are 
dominated by the possibility to have the tip of a jet within one bin. 
At asymptotic energies and large $q$ we then get the result 
\cite{Dahlqvist:1989yc,Gustafson:1991ru,Ochs:1992gd,Ochs:1992tg}
\begin{equation} 
D_q \approx \frac{q}{q-1} \sqrt{\frac{6 \alpha_s}{\pi}}
\label{eq:Dq}
\end{equation}

\begin{figure}
\includegraphics*[bb=0 0 580 560, width=7cm]{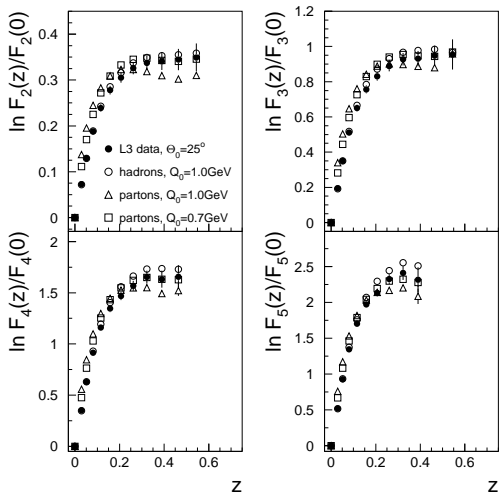}
\caption{\label{fig:intjetset}
L3 results for ratios of factorial moments in cones around a jet axis, 
$F_q(z)/F_q(0)$, together with JETSET 7.4 PS predictions on partonic
and hadronic levels. For details see ref.~\cite{Kittel:1997aw}.}
\end{figure}

\begin{figure}
\includegraphics*[bb=0 0 580 560, width=7cm]{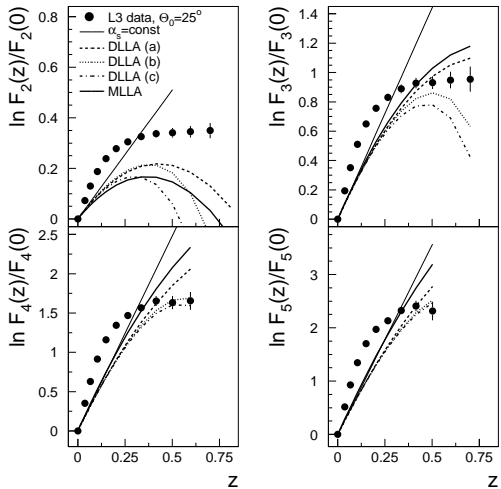}
\caption{\label{fig:intanalyt}
Analytic QCD predictions for $\Lambda=0.16$ GeV. The curves show results 
for $\alpha_s=$ const., 3 different DLL approximations and a MLL result. 
For details see ref.~\cite{Kittel:1997aw}.}
\end{figure}

As for the distributions in the full phase space, 
there are large corrections from non-leading 
effects and from hadronization. Therefore the analytic results 
have had limited success in comparisons with experimental data.
As examples figs.~\ref{fig:intjetset} and \ref{fig:intanalyt} 
show results from the L3 collaboration at LEP
\cite{Kittel:1997aw} for factorial moments in regions separated by
cones with opening angels $\Theta_0- \Theta$ and $\Theta_0+ \Theta$
around a jet axis. The variable $z$ is proportional to $\ln(\Theta_0/ \Theta$).
 We see that although MC programs work 
quite well (fig.~\ref{fig:intjetset}), the analytic calculations 
do not reproduce the data (fig.~\ref{fig:intanalyt}).
We note that different NLL calculations give quite different results.
These calculations have different approximations for NNLL
contributions, which shows that the higher order corrections cannot be
neglected. (It was early realised that BE 
correlations give an important contribution to the multiplicity moments,
and in the L3 analysis this effect is removed from the data.)

It is also possible to define the $\lambda$-measure for 
a hadronic state instead of a parton state, and it appears that analytic 
results for the $\lambda$-measure are closer to the
corresponding MC-results \cite{Gustafson:1991ru}. No comparisons with 
experimental data have been presented until now for such an analysis.

\subsection{Summary}

The gluon self coupling implies a fast growth of multiplicity and large 
fluctuations.
In LLA we get the following asymptotic results:
\begin{eqnarray}
\langle N\rangle &\sim&  \exp(2\sqrt{\alpha_0 \ln s};\,\,\,\,\,\,\,\, 
\alpha_0 \equiv \frac{3\alpha_s(Q^2)}{2\pi} \ln Q^2 \nonumber \\
V &=& \frac{1}{3} \langle N\rangle^2;\,\,\,\,\,\,\,\,\,\,\,\,\,\,\,\,\,\,\,
\gamma_0\equiv \frac{d \ln \langle N\rangle}{d \ln s} = 
\sqrt{\frac{3\alpha_s}{2\pi}}.
\end{eqnarray}

The QCD cascade has a fractal structure. The multiplicity in small 
phase space intervals shows intermittent features with a multifractal
(Renyi) dimension $D_q \sim 2 \gamma_0^{(s)}$, and the curve with the
$\lambda$-measure has the dimension $1+\gamma_0^{(s)}$.

However, energy-momentum conservation and other non-leading effects
are very important. NLL calculations give very different results
if they differ at NNLL order.
Hard emissions are the most important, but
matrix elements factorize only for soft emissions.
Therefore analytic calculations have in many cases met limited success. MC
programs work, however, generally quite well.

\section{Fermi-Dirac correlations in $pp$ and $\Lambda\Lambda$ pairs}

\subsection{Momentum correlations}

The results on baryon correlations presented here are obtained in 
collaboration with
R.M. Dur\'an Delgado and L. L\"onnblad \cite{rosa}.

Fermion correlation functions are usually fitted to the form
\begin{equation}
C(Q) = N \{1-\lambda e^{-R^2Q^2}\}.
\label{eq:corrfcn}
\end{equation}
The quantity $R$ is here interpreted as the radius of the production
region for the particle pair. The different LEP experiments have measured 
momentum correlations between $\Lambda\Lambda$ and/or
$\bar{p}\bar{p}$ pairs, and obtained production radii around 
$0.15\mathrm{fm}$, 
which is much smaller than the proton radius \cite{7, opal, aleph, delphi}.

In experiments the correlation function $C(Q)$ is determined by comparing
the observed number of pairs, $N(Q)$, with a reference sample, $N_{\mathrm{ref}}(Q)$:
\begin{equation}
C(Q)=\frac{N(Q)}{N_{\mathrm{ref}}(Q)}.
\end{equation}
The important question is how to construct the reference sample. Different 
methods have been applied:
\begin{itemize}
\item \emph{MC generation}. This is model dependent.
\item \emph{Mixed events}. The event structure is changed by gluon 
emission, which makes
it difficult to construct a sample of events with similar structure.
\end{itemize}
A common method to reduce the problems is to take a \emph{double ratio}:
\begin{equation}
C(Q)=\frac{N(Q)}{N_{\mathrm{mix}}(Q)} \bigg/ 
\frac{MC(Q)}{MC_{\mathrm{mix}}(Q)}.
\end{equation}
Here the sample of Monte Carlo events, $MC(Q)$, should be generated
without including FD correlations in the model. It is argued that the 
effect of different event types should be similar in the real data and
the MC events, and therefore be reduced in the ratio.

The mixed pairs depend only on inclusive spectra, and the MC
programs are tuned to reproduce the inclusive spectra with good precision.
This implies that the tuning also gives
\begin{equation}
N_{\mathrm{mix}}(Q) \approx MC_{\mathrm{mix}}(Q)
\end{equation}
which implies that
\begin{equation}
C(Q)\approx \frac{N(Q)}{MC(Q)}.
\end{equation}
Consequently the result \emph{depends very critically on a realistic MC},
also if the double ratio is used in the analysis.

\begin{figure}
 \begin{center}
  \includegraphics[angle=270, scale=0.28]{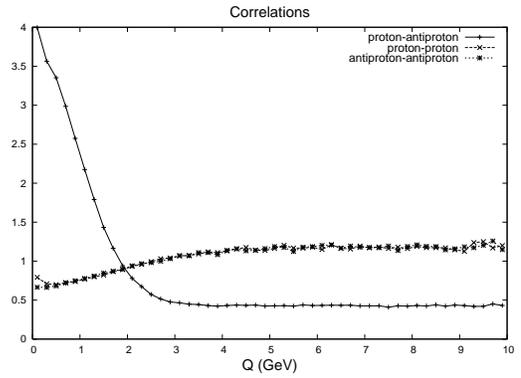}
\caption{\label{fig:pp-par} MC results for the ratio
$C_{MC}(Q) = MC(Q)/MC_{\mathrm{mix}}(Q)$ for
$pp$- ($\times$) and $p\bar{p}$-pairs (+).}
\end{center}
\end{figure}

In the Lund string hadronization model the colour string normally
breaks by $q\bar{q}$ pair production. The ordering of the hadrons
along the string, the rank ordering, agrees on average with the ordering 
in rapidity, with an average separation, $\Delta y$, of the order of half 
a unit in rapidity. A baryon-antibaryon pair can be produced
when a breakup is generated by a diquark-antidiquark pair, with
quark-antiquark pairs on either side. In this
picture two baryons must always be separated by at least one antibaryon
(and normally also with one or more mesons). This will give a strong
anti-correlation between two baryons in rapidity and in momentum.
This is seen in 
fig.~\ref{fig:pp-par}, which shows correlations in $pp$ and $p\bar{p}$
pairs. We see that there is a positive correlation between protons and 
antiprotons, which are neighbours in rank, but a negative correlation between two protons.

In the model there is also a correlation between momentum and space 
coordinates for the produced hadrons. Thus two identical baryons are 
(in the model) well
separated also in coordinate space, and we would from this picture
expect Fermi-Dirac correlations
to correspond to a radius $\sim 2 - 3$ fm. 

We see that the range for the $pp$ correlation in fig.~\ref{fig:pp-par}
is given by $Q \sim 1.5\mathrm{GeV} \sim 1/(0.15\mathrm{fm})$. We 
note that this corresponds exactly to the correlation length reported in
the experiments, although in this model the production radius is very 
much larger. The strength of the correlation is, however, smaller in 
the MC than in the data.

\emph{This raises the question: Is the difference between data and MC really a
FD effect, or could the MC underestimate the 
strength of the correlation?}
\vspace{1mm}

\begin{figure}
 \begin{center}
  \includegraphics[angle=270, scale=0.28]{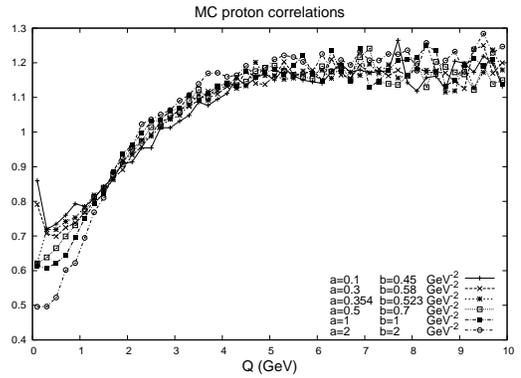}
\caption{\label{fig:abdep}
The ratio $C_{MC}(Q)= MC(Q)/MC_{\mathrm{mix}}(Q)$ for 
different values of the parameters $a$ and $b$. Larger $(a,b)$-values
give a stronger correlation and a deeper dip for small $Q$.}
\end{center}
\end{figure}

There are a number of sources for uncertainty in the MC:
\vspace{1mm}

1) There are two fundamental parameters, $a$ and $b$, in the Lund model
``splitting function''
\begin{equation}
f(z) \propto (-z)^a e^{-b m^2/z}
\label{eq:f(z)}
\end{equation}
The hadron multiplicity depends essentially on the ratio $(a+1)/b$,
This ratio is therefore well determined by experiments, but 
$a$ and $b$ separately are more uncertain. Small values of $a$ and $b$
correspond to a wide distribution $f(z)$, and a wide distribution in
the separation, $\Delta y$, between hadrons which are neighbours in rank.
Large values of $a$ and $b$ imply a narrow $\Delta y$- distribution
and therefore lower probability for two particles to be close in 
momentum space. The effect of varying $a$ and $b$ keeping the 
multiplicity unchanged is shown in fig.~\ref{fig:abdep}.

\begin{figure}
 \begin{center}
  \includegraphics[angle=270, scale=0.28]{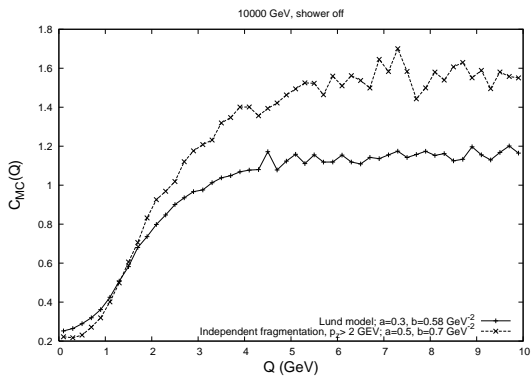}
  \caption{\label{fig:junc}
A single jet \emph{without} a junction has a large dip in 
$C_{MC}(Q)= MC(Q)/MC_{\mathrm{mix}}(Q)$ for small $Q$ ($\times$). 
This correlation is reduced by the approximate treatment in 
the MC of the small mass systems close to the "junction" (+).}
\end{center}
\end{figure}
\begin{figure}
 \begin{center}
  \includegraphics[angle=270, scale=0.28]{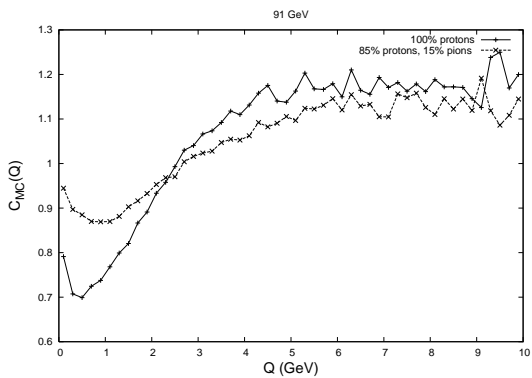}
  \caption{\label{fig:pion}
  $C_{MC}(Q)= MC(Q)/MC_{\mathrm{mix}}(Q)$ with (+) and without ($\times$)
  a 15\% admixture of pions.}
\end{center}
\end{figure}

2) The parameter $b$ is a universal constant, but $a$ may be different 
for baryons, although data are well fitted by a universal $a$-value. 
This gives some extra uncertainty.

3) The MC does not exactly reproduce the Lund hadronization model.
The splitting function in eq.~(\ref{eq:f(z)}) gives a correct result when
the remaining energy in the system is large. To minimise the error
at the end of the cascade, when the energy is small, the MC cuts off 
hadrons from both ends randomly, and joins the two ends when the 
remaining mass is small. This works well for inclusive distributions,
but implies that the correlations do not correspond to the model prediction for
particles close to the ``junction''. Fig.~\ref{fig:junc} shows results
for a single jet without a junction compared with the standard result.

4) Gluon emissions imply that straight string pieces are small compared to
the mass of a $B\bar{B}B$ system. This gives also extra uncertainty.

5) Pion correlations are not perfectly reproduced by the MC.
As an example the D\textsc{ELPHI} antiproton sample contains 15\% pions. 
Fig.~\ref{fig:pion} shows results with and without a 15\% pion admixture. 
We see that an error in the simulation of the pion-pion
or pion-proton correlations also affects the estimated proton-proton
correlations.

In summary we see that there are many effects which make the
MC predictions for $pp$ or $\Lambda\Lambda$ correlations quite uncertain.
As the experimental determination of the correlations rely so
strongly on a correct MC, it is therfore at present premature to
conclude that the production radius has the very small value around 0.15fm.

\vspace{3mm}
\subsection{Spin-spin correlations}

$\Lambda$ particles reveal their spin in the orientation of their decay 
products.
This has been used to study $\Lambda\Lambda$ correlations without the need for 
a comparison with MC results. A $\Lambda\Lambda$ pair with total spin 1 must
have an antisymmetric spacial wave function and is therefore expected to show 
a suppression for small relative momenta $Q$. $\Lambda\Lambda$ pairs with
total spin 0 has a symmetric spacial wave function, and should therefore
show an enhancement for small $Q$, similar to the correlation for bosons.
Therefore one expects pairs with small $Q$-values to be dominantly $S=0$.
Analyses at LEP \cite{5, 6, 7}  do indicate such an effect. They
are however based on rather low statistics, and the errors are presently
too large for any definite conclusion about the strength and range of the 
effect.

\subsection{Summary}

A production radius $R \sim 0.15$fm for baryon pairs is not consistent with
the conventional picture of string fragmentation.

Experimental results for $pp$ and $\Lambda\Lambda$ correlations depend 
sensitively on a reliable MC. 

The observed (anti-)correlation has the same range in $Q$ as the correlation 
in the MC, but is stronger.

Uncertainties in the MC implementation are large.

\emph{Conclusion}: It is therefore premature to claim evidence for a new 
production mechanism.

\bibliographystyle{utcaps}

\begin{thebibliography}{99}

\bibitem{Andersson:1988ee}
B. Andersson, P. Dahlqvist, and G. Gustafson,
{\em Phys. Lett.} {\bf B214} (1988) 604-608.

\bibitem{Gustafson:1993dd}
G. Gustafson and M. Olsson,
{\em Nucl. Phys.} {\bf B406} (1993) 293-324.

\bibitem{Andersson:1989ww}
B. Andersson, P. Dahlqvist, and G. Gustafson,
{\em Z. Phys.} {\bf C44} (1989) 455.

\bibitem{Andersson:1985qr}
B. Andersson, G. Gustafson, and B. S\"oderberg,
{\em Nucl. Phys.} {\bf B264} (1986) 29.

\bibitem{Dahlqvist:1989yc}
P. Dahlqvist, B. Andersson, and G. Gustafson, 
{\em Nucl. Phys.} {\bf B328} (1989) 76. 

\bibitem{Gustafson:1991ru}
G. Gustafson and A. Nilsson,
{\em Z. Phys.} {\bf C52} (1991) 533.

\bibitem{Ochs:1992gd}    
W. Ochs and J. Wosiek,
{\em Phys. Lett.} {\bf B289} (1992) 159. 

\bibitem{Ochs:1992tg}    
W. Ochs and J. Wosiek,
{\em Phys. Lett.} {\bf B305} (1993) 144.
   
\bibitem{Kittel:1997aw}
W. Kittel, S.V. Chekanov, D.J. Mangeol, and W.J. Metzger,
{\em Nucl. Phys. Proc. Suppl.} {\bf 71} (1999) 90;
hep-ex/9712003.

\bibitem{rosa}
R.M. Dur\'an Delgado, Diploma thesis,Lund preprint LU TP 06-17 (2006), and
R.M. Dur\'an Delgado, G. Gustafson, and L.~L\"onnblad, in preparation. 

\bibitem{7}
R. Barate et al. (ALEPH Coll.),
Phys. Lett. B475 (2000) 395.

\bibitem{opal}
OPAL Coll., CERN-EP/2001, OPAL-PN486 (2001).

\bibitem{aleph}
R. Barate et al., (ALEPH Coll.),
Phys. Lett. B611 (2005) 66-80

\bibitem{delphi}
DELPHI Coll., 
DELPHI 2004-038 CONF-713 (23 June, 2004), DELPHI 2005-010 CONF-730 (30 May, 2005).

\bibitem{5}
G. Alexander et al. (OPAL Coll.),
Phys. Lett. B384 (1996) 377.

\bibitem{6}
T. Lesiak and H. Palka (DELPHI Coll.),
CERN-EP/98-114, \\ DELPHI-CONF-176 (1998)






\end{thebibliography}


\end{document}